\documentclass[twocolumn,showpacs,superscriptaddress,preprintnumbers,aps,prl,10pt]{revtex4-1}
\usepackage{url}
\usepackage{dcolumn}
\begin{document}
\title{Comment on Phys.\ Rev.\ Lett.\  {\bf 109}, 152005 (2012)}
\author{A. Harindranath}
\affiliation{Theory Division, Saha Institute of Nuclear Physics,
1/AF Bidhan Nagar, Kolkata 700064, India}
\author{Rajen Kundu}
\affiliation{Department of Physics, RKMVC College, 
Rahara, Kolkata,  700118, West Bengal, India }
\author{Asmita Mukherjee}
\affiliation{Department of Physics, Indian Institute of Technology Bombay, Powai, 
Mumbai 400076, India}
\author{Raghunath Ratabole}
\affiliation{Department of Physics, BITS Pilani K K Birla Goa Campus,
NH17B, Zuarinagar, Goa 403726,  India}
\date{5 April 2013}

\begin{abstract}
 The identifications of transverse boost and rotation 
operators in light front theory done in Phys.\ Rev.\ Lett.\  {\bf 109}, 152005 
(2012) is incorrect. The simple parton interpretation claimed is, in fact,
for the transverse boost operator. Manifestation of Lorentz symmetry as claimed in the 
context of their calculation involving transverse Pauli-Lubanski polarization 
vector is unsupported. 
\end{abstract}

\maketitle

The identifications of  the transverse rotation and the transverse
boost operators  done in Ref. \cite{jxy} following Ref. \cite{jxyplb} by the
same authors 
is incorrect, as also pointed out in  Ref. \cite{ll}.
The correct identifications  and the 
associated sum rules  in light 
front QCD were investigated by us \cite{hk,hmr1,hmr2} a decade ago.
In the following we elaborate on this.

 According to Ref. \cite{jxyplb}, $``$ In order to obtain the
boost-invariant spin sum rule $\ldots$ we construct the polarization through
the Lorentz-covariant  Pauli-Lubanski vector". (Note that, 
the terms  spin and polarization are used interchangeably throughout this
paper and  the difference between the two, if
any, is never really clarified.) However, the transverse components of the
Pauli-Lubanski operator $W^i$'s ($i=1,2$) are {\em not boost invariant} in 
light front dynamics (for a review, see Ref. \cite{bpp}) whereas 
the intrinsic spin operators ${\cal J}^i$ are \cite{soper,ls}. The two
are related by $M{\cal J}^i = W^i - P^i {\cal J}^3$ and are same only in the 
$P^i=0$ frame upto a constant factor. In our works in 
Refs. \cite{hk,hmr1,hmr2}, we start 
from ${\cal J}^i$ and naturally arrive at frame independent results. In 
Refs. \cite{jxy,jxyplb}, 
they start from $W^i$ and their subsequent results and conclusions, if at all 
valid  (see additional comments below), hold only in $P^i=0$ frame, contrary 
to their claim.        

In Ref. \cite{jxyplb}, after Eq. (9), $J^{+\sigma}$ is
identified as angular momentum operator and $J^{-\sigma}$ is identified as
boost operator. {\em This is wrong}. For $\sigma=\perp$ which are the
relevant components under discussion, it is well-known that $J^{+\sigma}$
which are kinematical are the transverse boost operators and
$J^{-\sigma}$ which are dynamical are the transverse rotation 
operators. The simple parton interpretation claimed is, in fact,
for the transverse boost operator in Eq. (2) in Ref. \cite{jxy}.



Contrary to the statement made in Ref. \cite{jxyplb} that $``$we take no
contribution to $W^\perp_i$ from the energy momentum tensor $T^{+-}$", we
find \cite{hkm} that (i) both the form factors $A_i$ and ${\bar C}_i$ 
contribute
to the matrix element of $T_i^{+-}$ in a transversely polarized state, (ii) 
there is no relative suppression factor between these two contributions and 
(iii) 
the contribution to $W_i^\perp$  from $T_i^{++}$ contains only the form 
factor 
$B_i$ and not the form factor $A_i$. (Incidentally, the last finding is 
already a  well established result \cite{bhms}.) 

Thus, we conclude that in Ref. \cite{jxyplb}, (i) there is no justification 
for ignoring the contribution of ${\bar C}_i$ to $W^\perp_i$ as has been done, 
(ii) the claim in Eq. (29) 
is unsupported and (iii) so are the claims 
made after Eq. (30) that $``$ $T^{++}_i$ and $T^{+\perp}_i$ contribute
separately 1/2 of the nucleon spin" and $``$This is a simple result of 
Lorentz symmetry". In fact, borrowing one of their argument for dropping 
${\bar C}_i$,  it follows that since $B$ form factor does
not contribute to transverse spin sum rules, (as $B_q+B_g=0$ where $q$ and 
$g$ denote
quark and gluon parts), matrix element of $T^{++}$  does not contribute at all
contrary to the claim in 
Eq. (29). Moreover, if the higher twist contribution is replaced by leading 
twist contribution as they claim due to Lorentz symmetry, the distiction 
between leading and subleading contributions are washed away.
Lastly, based on the extra factor of $P^+$ in Eq. (2) for transverse boost
matrix element, compared to Eq. (3) for the matrix element of helicity,  
Ref. \cite{jxy} claims that nucleon helicity is
a sub-leading quantity whereas transverse polarization is a leading
quantity. This claim has no basis.


\end{document}